\documentclass[twocolumn,showpacs,aps,prl,amsmath,amssymb,nofootinbib,nobibnotes,floatfix]{revtex4-1}

\usepackage{hyperref}
\usepackage{amsfonts}
\usepackage{amsmath}
\usepackage{mathrsfs}
\usepackage{epsfig}
\usepackage{graphicx}               
\usepackage{url}
\usepackage{hyperref}
\usepackage{float}
\usepackage{color}
\usepackage[utf8]{inputenc}

\usepackage{graphicx}
\usepackage{epsf}

\usepackage{comment}

\newcommand{\nc}{\newcommand}

\nc{\beq}{\begin{equation}}
\nc{\eeq}{\end{equation}}
\nc{\beqa}{\begin{eqnarray}}
\nc{\eeqa}{\end{eqnarray}}

\usepackage{slashed}

\newcommand{\lsim}{\!\mathrel{\hbox{\rlap{\lower.55ex \hbox{$\sim$}} \kern-.34em \raise.4ex \hbox{$<$}}}}
\newcommand{\gsim}{\!\mathrel{\hbox{\rlap{\lower.55ex \hbox{$\sim$}} \kern-.34em \raise.4ex \hbox{$>$}}}}

\def\be{\begin{equation}}
\def\ee{\end{equation}}

\newcommand\affspc{\vspace{4pt}}

\usepackage{footmisc}
\usepackage{setspace}
\begin{document}

\title{Superradiant Instability and Backreaction of Massive Vector Fields around Kerr Black Holes}

\author{William E.\ East$^1$ and Frans Pretorius$^2$}
\affiliation{$^1$Perimeter Institute for Theoretical Physics, Waterloo, Ontario N2L 2Y5, Canada \affspc}
\affiliation{$^2$Department of Physics, Princeton University, Princeton, New Jersey 08544, USA \affspc} 

\begin{abstract}
We study the growth and saturation of the superradiant instability of a complex, massive
vector (Proca) field as it extracts energy and angular
momentum from a spinning black hole, using numerical solutions of the full
Einstein-Proca equations.  We concentrate on a rapidly spinning black hole
($a=0.99$) and the dominant $m=1$ azimuthal mode of the Proca field, with
real and imaginary components of the field chosen to yield an axisymmetric
stress-energy tensor and, hence, spacetime. We find that in excess of $9\%$ of
the black hole's mass can be transferred into the field.  In all cases
studied, the superradiant instability smoothly saturates when the black hole's
horizon frequency decreases to match the frequency of the Proca cloud that
spontaneously forms around the black hole. 
\end{abstract}

\maketitle

{\em Introduction.}---%
A remarkable feature of spinning black holes (BHs) is that a portion of their
mass---up to $29\%$ for extremal spin---can, in principle, be
extracted.  One way to realize this liberation of rotational energy is through
the interaction of the BH with an impinging wave---be it scalar,
electromagnetic, or gravitational---with frequency $\omega<m\Omega_{\rm BH}$,
where $\Omega_{\rm BH}$ is the BH horizon frequency and $m$ is the azimuthal
number of the wave.  Waves satisfying this criterion exhibit superradiance and
carry away energy and angular momentum from the BH.
An analogous phenomenon can occur for charged BHs,
where the electromagnetic energy of the BH is
superradiantly transferred to an interacting charged matter field interacting with the BH.

Going back to Ref.~\cite{1972Natur.238..211P}, there has been speculation of how
superradiance could be combined with a confining mechanism to force the wave to
continuously interact with the BH and hence undergo exponential growth---a so
called ``black hole bomb."
The first nonlinear studies of this process were recently undertaken for a charged
scalar field around a charged BH in spherical symmetry, both in a reflective
cavity in asymptotically flat space~\cite{Sanchis-Gual:2015lje}, and in the naturally
confining environment of an asymptotically anti-de Sitter domain~\cite{Bosch:2016vcp}. 

However, there is an exciting possibility that a variation of this scenario
could, in fact, be realized around astrophysical spinning BHs. Massive bosonic fields
with a Compton wavelength comparable to, or larger than, the horizon radius of a BH can
form bound states around the BH, and if the latter is spinning, the bound states
can grow from a seed perturbation through superradiance~\cite{Damour:1976,Detweiler:1980uk,Zouros:1979iw}.
This implies that stellar mass BHs can probe the existence of ultralight bosons with masses
$\lesssim10^{-10}$ eV that are weakly coupled to ordinary matter and thus difficult to detect by other
means. Theoretical scenarios where this might occur include the string
axiverse~\cite{Arvanitaki:2009fg,Arvanitaki:2010sy}, the QCD
axion~\cite{Arvanitaki:2014wva}, and dark photons~\cite{Holdom:1985ag,Pani:2012bp}.
Such particles could form large clouds, spinning down the BH
in the process. This is of particular interest now that LIGO has begun
observing gravitational waves (GWs)~\cite{Abbott:2016blz}, since measurements of
BH masses and spins from binary mergers can be used to rule out or provide
evidence for such particles, in addition to direct searches for the GW
signatures of boson clouds~\cite{Arvanitaki:2016qwi,BaryakhtarNew}. 
See~\cite{2015LNP...906.....B} for a review.

Though details of the nonlinear growth and saturation of the rotational superradiant instability
will be important to help observe or rule out such massive fields, there
are presently few results of relevance to this regime where the backreaction on the 
BH is significant. In Ref.~\cite{East:2013mfa}, it was found that, for sufficiently
large GWs superradiantly scattering off a Kerr BH, 
backreaction effects decrease the efficiency of energy extraction
(for the analogous case of the scattering of a charged scalar
field by a Reissner-Nordstr\"om BH, see~\cite{Baake:2016oku}).
The nonlinear behavior of the superradiant instability of massive bosons has not
been addressed before. This is because of the computational cost of solving the
equations, in part due to the disparate time scales between the oscillation of the field
and the growth rate of the instability and the lack of 
symmetries to reduce it to a $(1+1)$-dimensional problem (unlike the charged case).
Important questions include what the
efficiency of energy and angular momentum extraction is, how explosive
the nonlinear phase of growth is (e.g., can the energy extraction
overshoot limits implied by the parameters of the field and BH~\cite{Sanchis-Gual:2015lje}), 
and what the final state is after a non-negligible amount of energy has been
transferred to the Proca field (e.g., does a stable cloud form around the BH,
or could there be something akin to a \emph{bosenova}
where the entire field is rapidly expelled from the vicinity of the BH).

In this Letter we begin to address these questions related to the nonlinear behavior of the
superradiant instability of massive bosonic fields around a spinning BH. We focus on the case of
a vector field, as it exhibits faster growth than a scalar field.
The linear regime of the instability for Proca fields has been studied before
in various limits~\cite{Pani:2012bp,Pani:2012vp,Endlich:2016jgc,Witek:2012tr,BaryakhtarNew}.
Here we find numerical solutions of the
full Einstein-Proca field equations.
To make the problem computationally tractable, we use a complex field with prescribed
$m=1$ azimuthal dependence to give an axisymmetric stress-energy tensor and, hence,
spacetime geometry.
Beginning with a
seed field about a rapidly rotating BH, we find that the instability
efficiently grows into the nonlinear regime and smoothly saturates when the BH
horizon frequency decreases to match that of the Proca cloud.
This frequency depends on the mass parameter of the field, and for a value
near where we expect maximal energy extraction, we find that when the instability
saturates a large Proca cloud has formed, containing $9\%$ of the initial BH mass
(and $38\%$ its initial angular momentum). 
We use units with $G=c=1$ throughout.

{\em Methodology.}---%
We consider a Kerr BH with initial mass $M_0$ and
dimensionless spin $a=0.99$ in the presence of a complex Proca field $X^a$
with constant mass parameter $\mu$, and numerically evolve the coupled Einstein-Proca equations.
The Proca field equation of motion is $\nabla_a F^{ab}= \mu^2 X^b$, where
$F_{ab}=\nabla_a X_b-\nabla_b X_a$, and its corresponding stress-energy tensor is  
\beqa
T_{ab}&=&\frac{1}{2}(F_{ac}\bar{F}_{bd}+\bar{F}_{ac}F_{bd})g^{cd}-\frac{1}{4}g_{ab}F_{cd}\bar{F}^{cd}
\nonumber \\ &&+\frac{\mu^2}{2}(X_a\bar{X}_b+\bar{X}_aX_b-g_{ab}X_c\bar{X}^c) ,
\eeqa 
where the overbar indicates complex conjugation.  
We evolve the Proca equations in a form similar to Ref.~\cite{Zilhao:2015tya}
(which also evolved the Einstein-Proca equations, though without symmetry
restrictions and focusing on nonlinear interactions between the field and a nonspinning BH).
We restrict 
to cases where the Proca field has an $m=1$ azimuthal dependence and the
resulting stress-energy tensor and spacetime are axisymmetric---i.e., in terms
of the Lie derivative with respect to the axisymmetric Killing vector $(\partial/\partial \phi)^b$,
$\mathcal{L}_{\phi}X_a=iX_a$.  This allows us to use a two-dimensional
numerical domain for the spatial discretization, which is essential in making evolutions on time scales of
$\sim10^5 M_0$ computationally feasible. Here we study cases with
$\tilde{\mu}:=M_0\mu=$0.25, 0.3, 0.4, and 0.5. As we discuss below,
$\tilde{\mu}=0.25$ is near the value that maximizes the energy extracted
from the BH, while $\tilde{\mu}=0.5$ is close to the value that gives the maximum growth rate for the
linear instability~\cite{proca_linear}.  We begin with a seed Proca field with
energy $\sim10^{-3}M_0$ around a Kerr BH (ignoring the effect of this small
field on the initial spacetime geometry) and study the subsequent evolution.  
We have verified that using different or lower amplitude perturbations gives similar
results.
 
We evolve the Einstein equations using the generalized harmonic formulation,
with the gauge degrees of freedom set by fixing the source functions to the
values of the initial BH solution in Kerr-Schild coordinates, as
in Ref.~\cite{East:2013mfa}.  To mitigate the accumulation of truncation error
during the long period it takes for the Proca field to grow large enough to
significantly backreact on the spacetime, we use
the background error subtraction technique described in Ref.~\cite{best}, with
the initial isolated spinning BH as the background solution.  As the spacetime evolves,
we keep track of the BH apparent horizon and measure its area $A$ and angular
momentum $J_{\rm BH}$, from which we can derive a mass using the Christodoulou formula
\beq
M_{\rm BH} = \left(M_{\rm irr}^2+\frac{J_{\rm BH}^2}{4M_{\rm irr}^2}\right)^{1/2} , 
\eeq
where $M_{\rm irr}=\sqrt{A/16\pi}$ is the irreducible mass. We
also measure the flux of Proca field energy $\dot{E}^{H}$ and angular momentum
$\dot{J}^H$ through the BH horizon.  In addition, for the Proca field we keep
track of the energy density $\rho_E=-\alpha T^t_t$ and angular momentum density
$\rho_J=-\alpha T^t_{\phi}$ (where $\alpha=[-g^{tt}]^{-1/2}$ is the lapse), the volume integrals
of which give a measure of the total energy $E$ and angular momentum $J$
outside of the BH horizon.  Details on the numerical resolution and convergence are
given in the appendix; more information on how we evolve the Proca equations,
as well as results on the instability in the test-field regime,
are provided in Ref.~\cite{proca_linear}.

\begin{figure}
\begin{center}
\includegraphics[width=\columnwidth,draft=false]{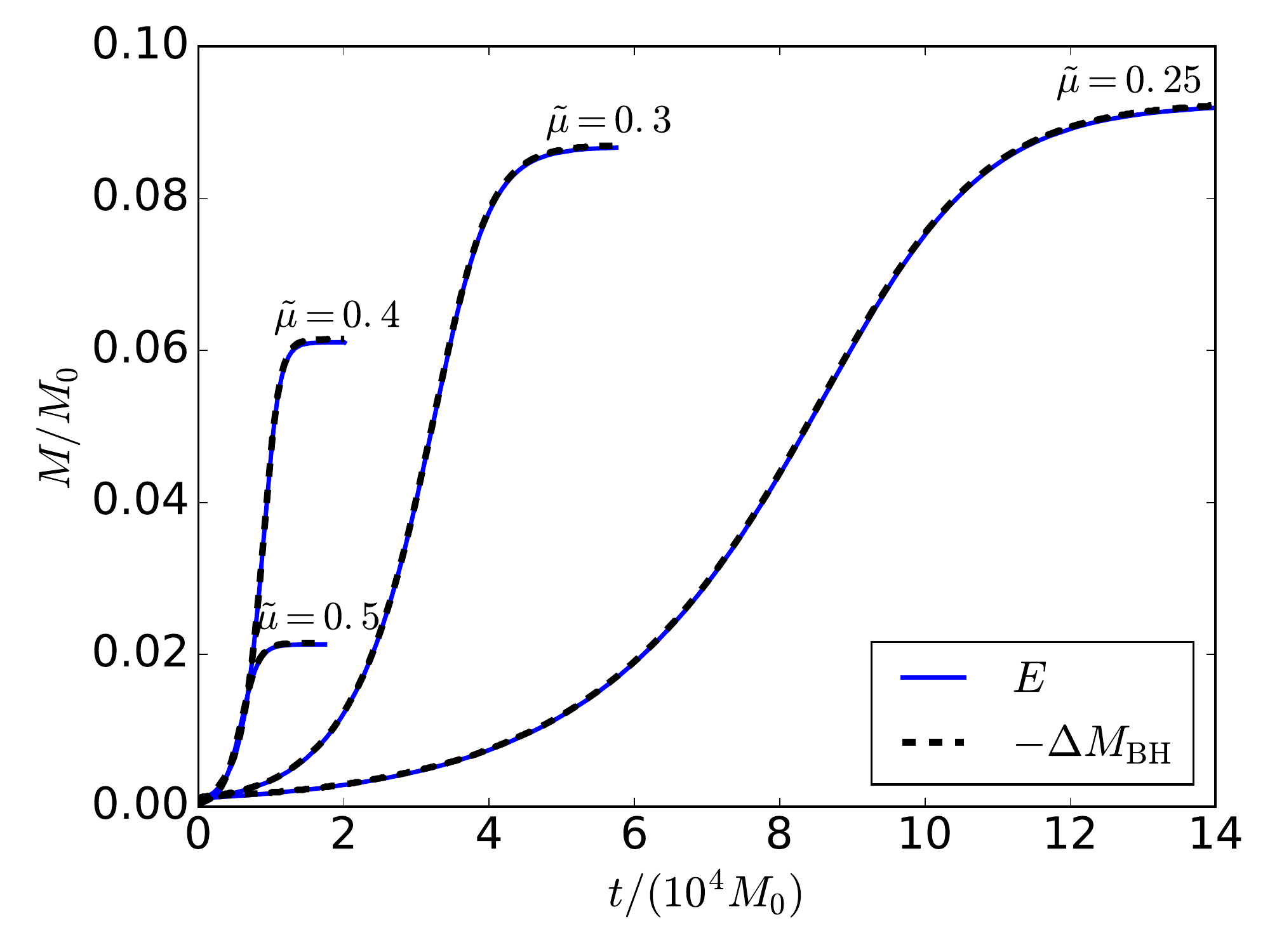}
\includegraphics[width=\columnwidth,draft=false]{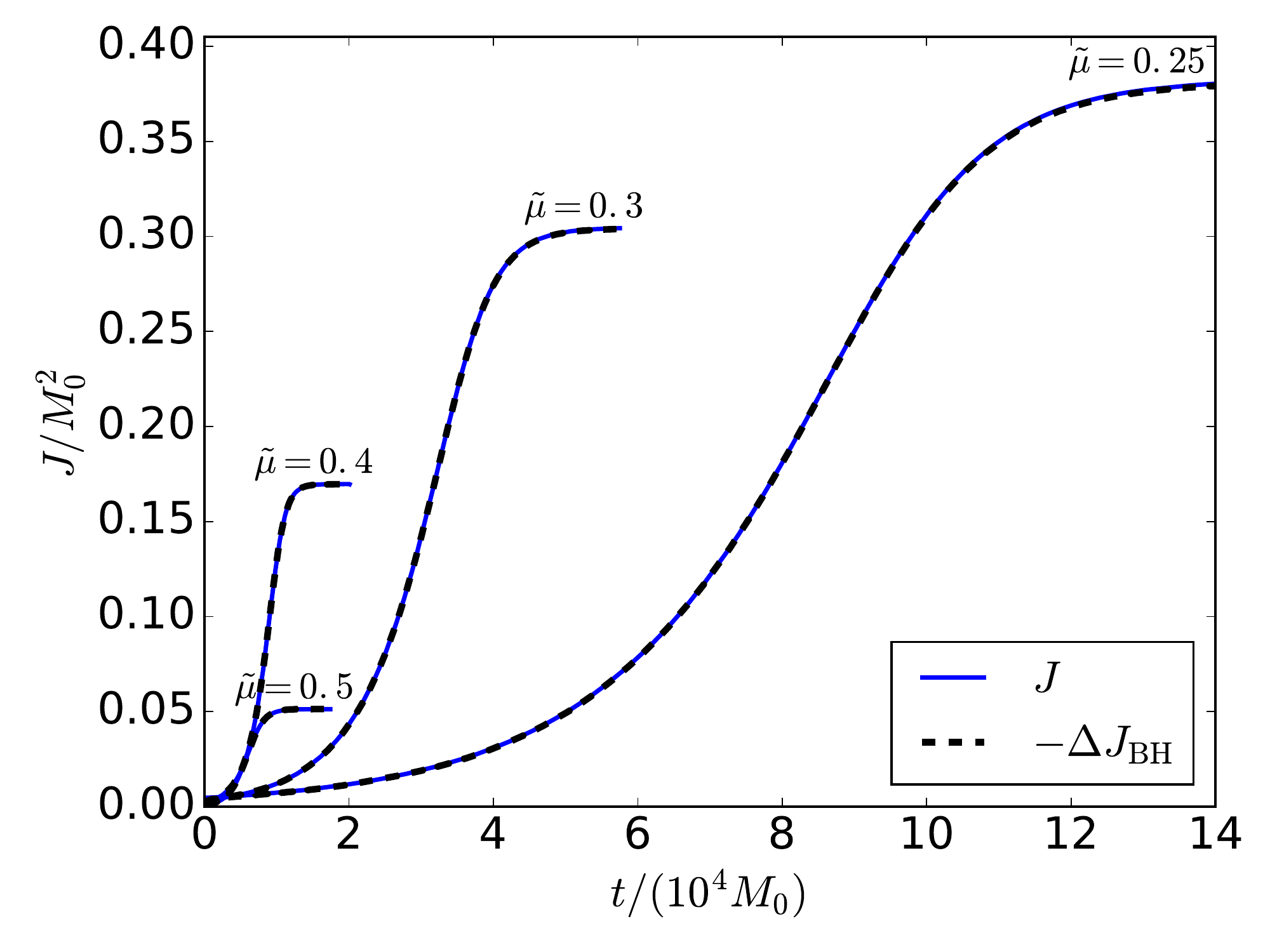}
\end{center}
\caption{
The energy (top) and angular momentum (bottom) in the Proca field
as a function of time (solid lines), along with the loss in mass (top) and angular momentum (bottom)
of the BH (dashed lines). 
\label{fig:ej}
}
\end{figure}

{\em Results.}---%
All the cases studied here are susceptible to a linear superradiant
instability, and after a brief transient period the energy and angular momentum
in the Proca field enter a period of exponential growth as shown in
Fig.~\ref{fig:ej}.  As also shown there, the corresponding loss
of mass and angular momentum by the BH, as measured from its horizon
properties, closely tracks this.  The cases with larger $\mu$ have larger
growth rates for the instability and also saturate with smaller energy and
angular momentum.  Though the mass of the BH is decreasing in each case, as
required by BH thermodynamics the irreducible mass $M_{\rm irr}$ is always
increasing, 
and smaller $\mu$ cases
saturate with a larger overall increase in $M_{\rm irr}$.

\begin{figure}
\begin{center}
\includegraphics[width=\columnwidth,draft=false]{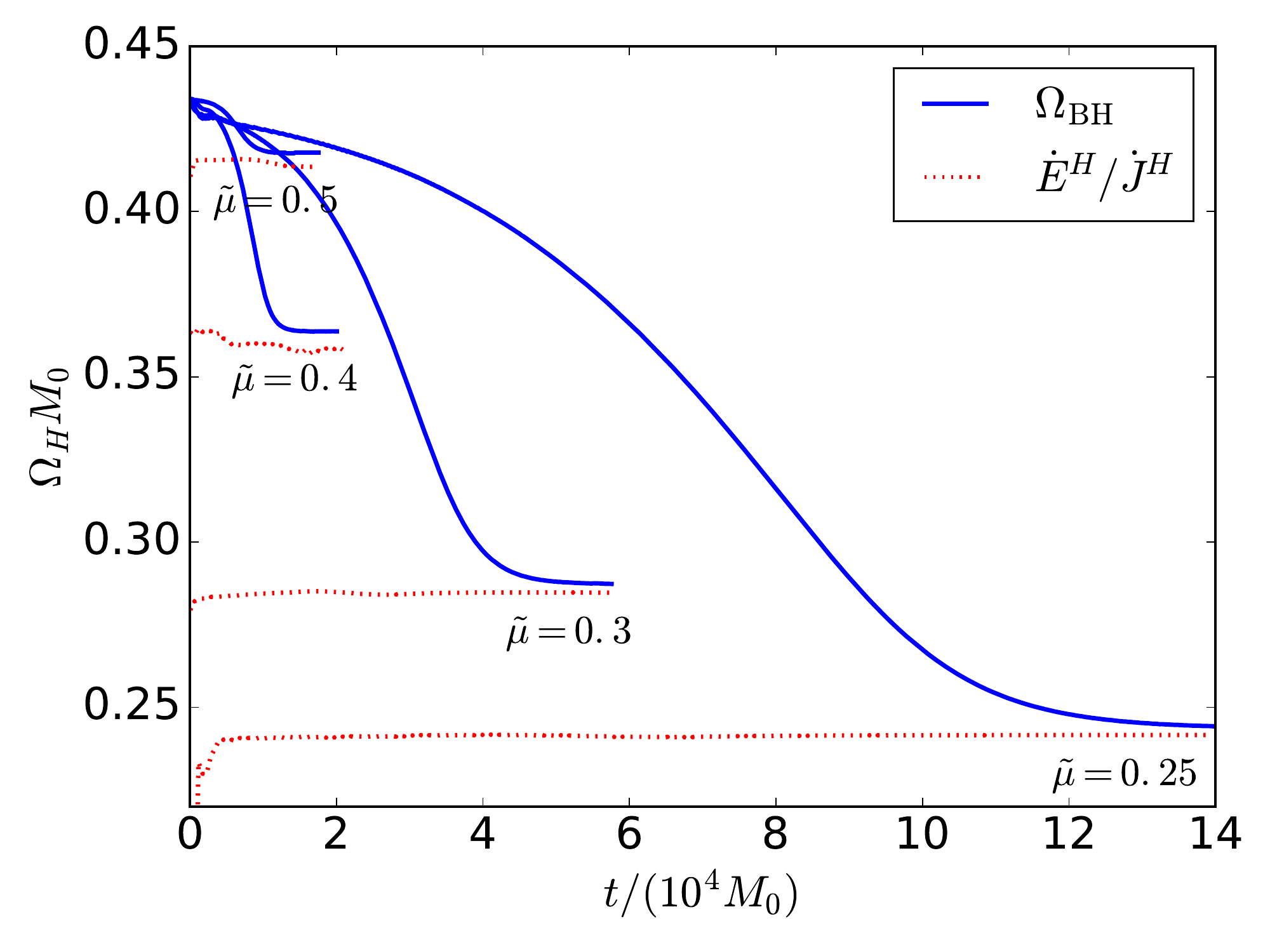}
\end{center}
\caption{
The BH horizon frequency $\Omega_{\rm BH}$, as calculated
from the BH's mass and angular momentum, and the ratio of the flux
of Proca field energy and angular momentum $\dot{E}^H/\dot{J}^H$ 
through the BH horizon, as a function of time.
\label{fig:omega}
}
\end{figure}

The reason for the saturation of the superradiant instability is illustrated in
Fig.~\ref{fig:omega}, where we plot both the horizon frequency of the BH 
$\Omega_{\rm BH}$ and the ratio of Proca field energy to angular momentum flux
through the horizon $\dot{E}^H/\dot{J}^H$.  When $\Omega_{\rm
BH}>\dot{E}^H/\dot{J}^H$, the superradiant condition is met and the Proca
cloud will extract rotational energy from the BH.  However, as
shown in Fig.~\ref{fig:omega}, eventually the BH's horizon
frequency decreases to the point where $\Omega_{\rm
BH}\approx\dot{E}^H/\dot{J}^H$, and the instability saturates.

\begin{figure}
\begin{center}
\includegraphics[width=\columnwidth,draft=false]{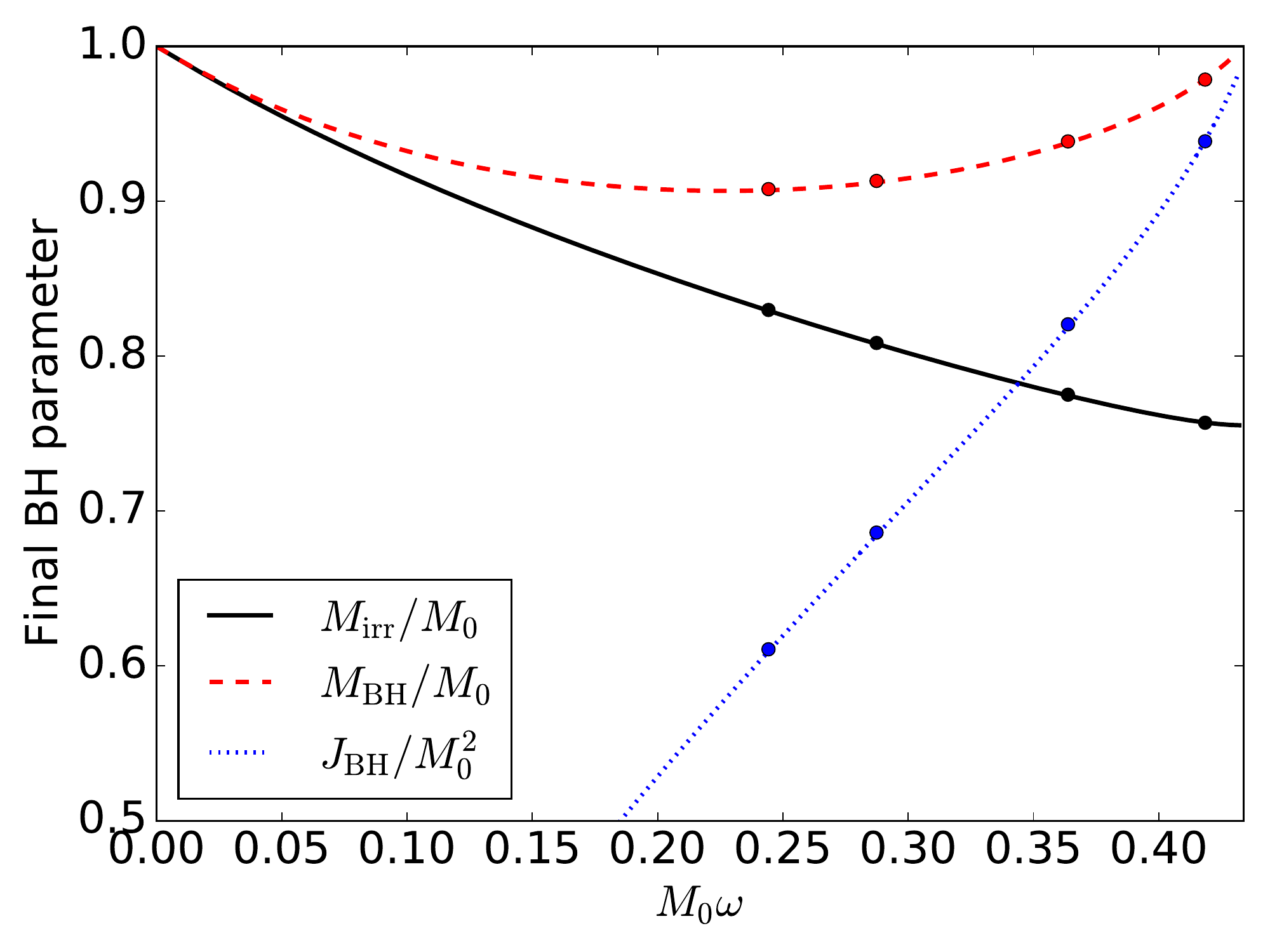}
\end{center}
\caption{
The final BH irreducible mass, total mass, and angular momentum after saturation of the
$m=1$ superradiant instability for a BH with $a=0.99$ ($M_{{\rm irr},0}/M_0\approx 0.76$) initially.  The lines
show the prediction obtained with the assumption that the BH will lose energy
and angular momentum in fixed proportion $\omega$, until $\Omega_{\rm
BH}=\omega$, while the points show the measured values from the simulations. 
\label{fig:fbh}
}
\end{figure}

We can obtain simple estimates of the final state properties of the black hole
if we assume, as roughly consistent with the simulations, that 
the instability will extract energy and angular momentum in some
fixed proportion $\omega(\mu)=\dot{E}^H/\dot{J}^H$ 
[where $\omega(\mu)\approx \mu(1-\tilde{\mu}^2/2)$
in the linear/small $\tilde{\mu}$ limit~\cite{Dolan:2007mj,Rosa:2011my}]
until
$\omega(\mu)=\Omega_{\rm BH}$. We plot the results in Fig.~\ref{fig:fbh}, 
along with the four end-state points from the full nonlinear simulations,
showing excellent agreement with the approximation. 
This indicates an efficient extraction of energy and
angular momentum, with a negligible additional increase in irreducible
mass (equivalently, BH entropy). This is likely due to the relatively
slow evolution of the instability compared to the light-crossing time
of the BH, even approaching saturation (similar conclusions
were reached using a ``quasiadiabatic'' approximation for the massive scalar field 
instability in Refs.~\cite{Dolan:2012yt,2015CQGra..32m4001B}).
We see that the energy lost by the BH should be maximized at $-\Delta M_{\rm
BH}/M_0\approx0.093$, 
near the value $-\Delta M_{\rm
BH}/M_0\approx0.092$ found for $\tilde{\mu}=0.25$ here.
For lower values of $\mu$, less energy, but
more angular momentum will be extracted, with the instability just converting
the Kerr BH into a nonspinning BH of the same mass in the $\mu \to 0$ limit.  

\begin{figure}
\begin{center}
\includegraphics[width=\columnwidth,draft=false]{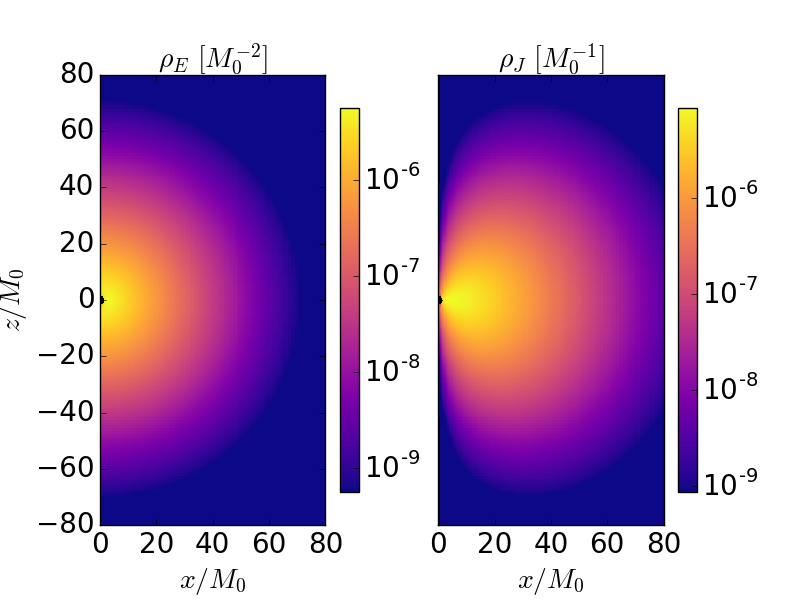}
\includegraphics[width=\columnwidth,draft=false]{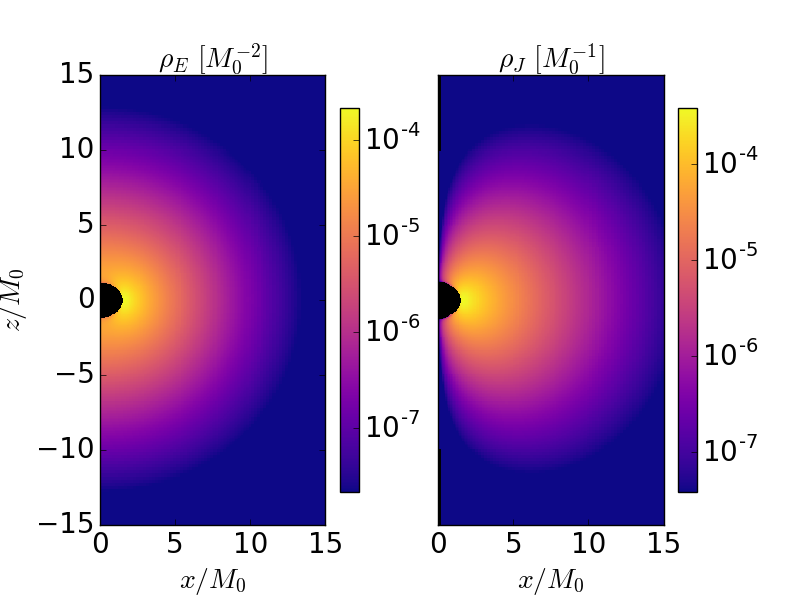}
\end{center}
\caption{
The energy (left) and angular momentum density (right) of the Proca field in the final state
with $\tilde{\mu}=0.25$ (top) and $\tilde{\mu}=0.5$ (bottom)
for a slice containing the
BH spin axis (the $z$ axis) and perpendicular to the equatorial plane ($z=0$); note the different scales
for the two cases. 
\label{fig:ejpr}
}
\end{figure}

After saturation, the resulting 
configuration consists of a BH surrounded by a Proca cloud with
roughly stationary energy density, though the phase of the complex field is oscillating 
at a constant frequency.
The energy and angular momentum density of the resulting clouds are illustrated
in Fig.~\ref{fig:ejpr} for two cases.  Away from the BH, the Proca clouds have a roughly
spherical energy density, falling off exponentially with distance from the BH.  
For the larger $\mu$ cases, the cloud is concentrated on much
smaller scales near the BH.

As expected, given the close match between the energy and angular momentum lost by
the BH and that gained by the Proca cloud, the radiation from both 
GWs and the Proca field is negligible (the dominant
contribution of which comes from the other modes in the seed perturbation
leading to initial radiation).
The fact that we are considering a complex
Proca field and restricting our study to axisymmetric spacetimes will suppress
the gravitational radiation. 
We can estimate what the
gravitational radiation would be for a single real Proca field by 
using the GW luminosity results from the test field limit~\cite{proca_linear} and scaling them
using $P_{\rm GW}\propto E^2$.
This gives 
$P_{\rm GW}\sim 6\times10^{-8}$, $2\times10^{-7}$, $6\times10^{-7}$, and $6\times10^{-8}$ 
for the Proca field clouds at the end of the
$\tilde{\mu}=0.25$, 0.3, 0.4, and 0.5 simulations, respectively.
This means that once the BH has spun down to below the
superradiant regime, the Proca cloud will decay via GW emission on time scales of $\sim10^5$--$10^6M_0$.

Our Proca field ansatz only allows exploration of the $m=1$ mode instability.
Higher $m$ modes are also unstable, even after the BH has spun down
to the point where the $m=1$ becomes stable (ignoring the Proca cloud).
However, the grow rates become significantly longer with increasing $m$.  For
example, for $\tilde{\mu}=0.5$ the $m=2$ instability has a growth rate
$\sim2000$ times as long as the $m=1$ mode~\cite{proca_linear}, and the disparity is worse for smaller
$\mu$ (with the relative growth rates scaling as $\mu^4$).

{\em Conclusion.}---%
We have studied the growth and saturation of the superradiant instability of a
massive vector field around a Kerr BH.  We find that in all cases the
instability efficiently extracts energy from the BH and then smoothly shuts off
as the BH horizon frequency decreases to the threshold of instability.  This
contrasts with Ref.~\cite{Sanchis-Gual:2015lje}, where the energy extraction of a
charged BH in a reflecting cavity was seen to overshoot in some cases; this
could be due to the presence of multiple unstable modes~\cite{Bosch:2016vcp}.
We further find that at saturation essentially all the energy and angular
momentum extracted from the BH has gone into forming a cloud of complex Proca
``hair" with stationary energy density surrounding the BH. A family of
stationary hairy BH solutions with this property and the same matter model was
constructed in Ref.~\cite{Herdeiro:2016tmi} which is plausibly the same as our end
states; it would be interesting to investigate in detail how close these
solutions are to what we find at saturation.  In our case, the Proca clouds
persist for the relatively short times we have extended the runs beyond
saturation, though this is not adequate to comment on their long-term
stability.

{\em Acknowledgements.}---
We thank Asimina Arvanitaki, Masha Baryakhtar, Stephen Green, Robert Lasenby,
Luis Lehner, Vasilis Paschalidis, Justin Ripley, Kent Yagi, and Huan Yang for
stimulating discussions.  This research was supported in part by Perimeter
Institute for Theoretical Physics (WE), the National Science Foundation through
Grant No. PHY-1607449 (FP), and the Simons Foundation (FP). Research at
Perimeter Institute is supported by the Government of Canada through the
Department of Innovation, Science and Economic Development Canada and by the
Province of Ontario through the Ministry of Research, Innovation and Science.
Simulations were run on the Perseus Cluster at Princeton University and the
Sherlock Cluster at Stanford University.

\bibliographystyle{h-physrev}
\bibliography{ref}

\appendix
\section{Details of numerical methods}
\begin{figure}
\begin{center}
\includegraphics[width=\columnwidth,draft=false]{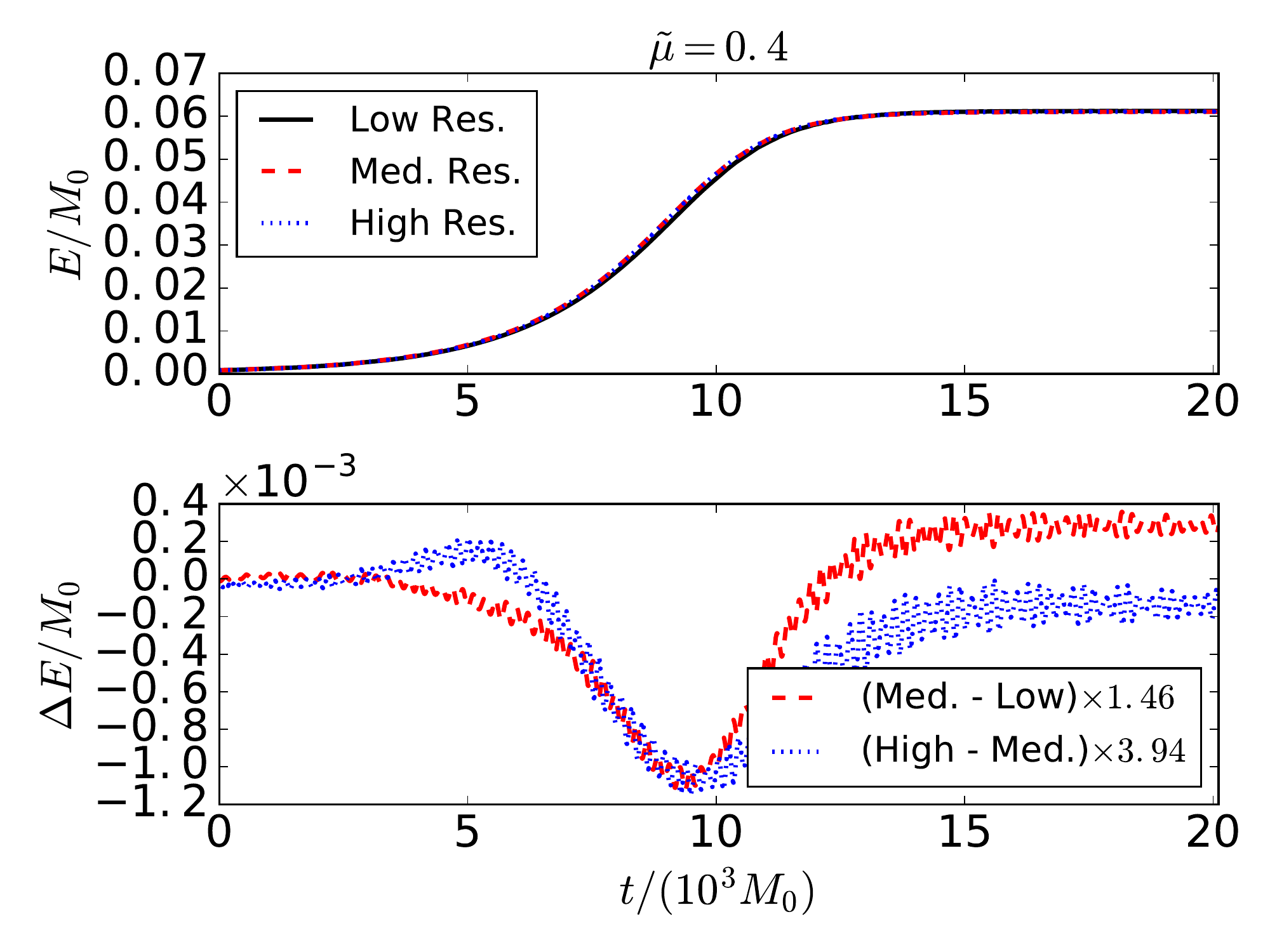}
\includegraphics[width=\columnwidth,draft=false]{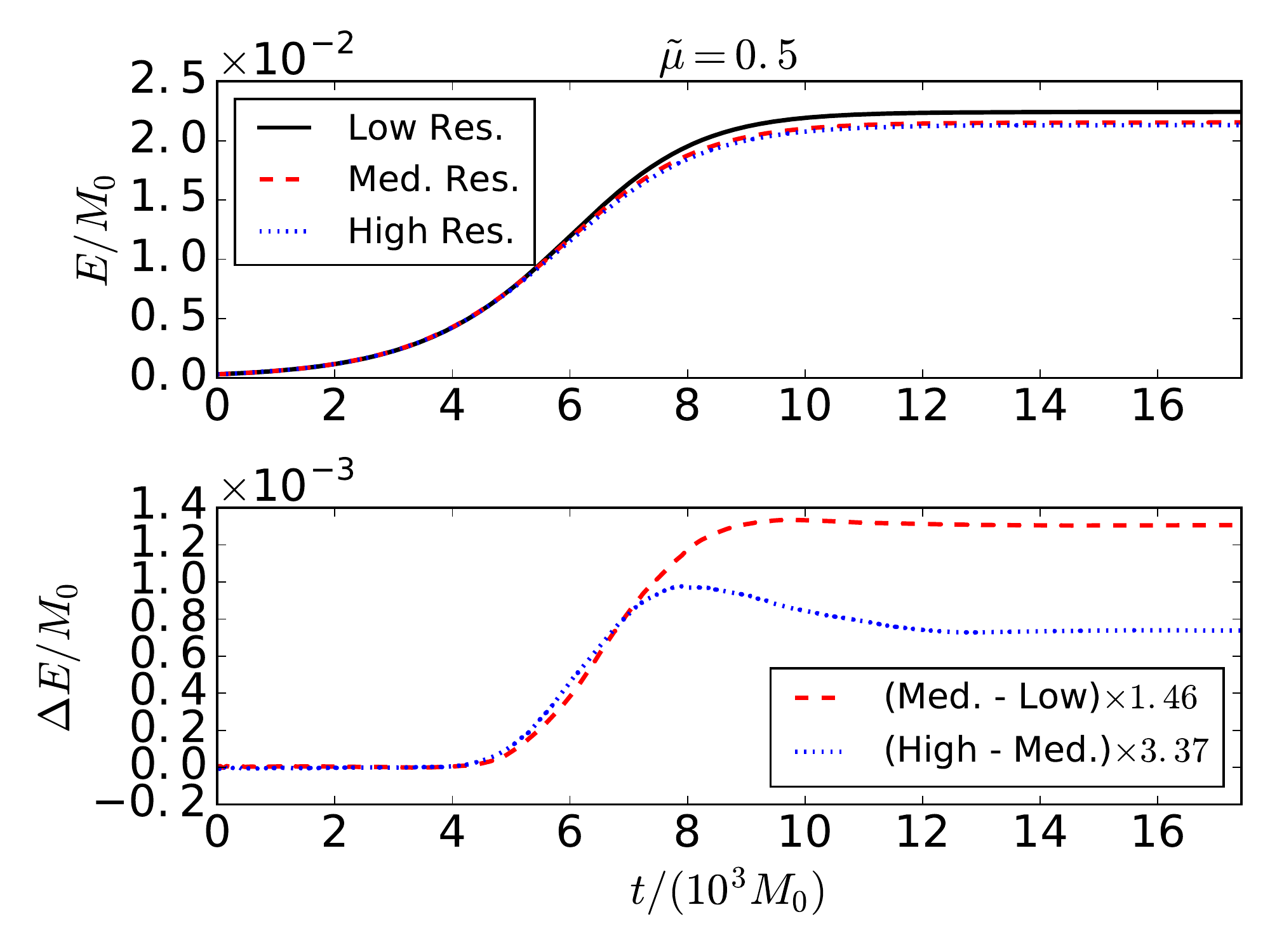}
\end{center}
\caption{
The energy in the Proca field for simulations performed at three different resolutions, and
the difference in this quantity with resolution, scaled assuming fourth-order convergence
for $\tilde{\mu}=0.4$ (top) and $\tilde{\mu}=0.5$ (bottom).
\label{fig:econv}
}
\end{figure}

We employ fourth-order accurate finite difference methods to integrate the
equations.  For all of the simulations presented here we used a grid hierarchy
with seven levels of mesh refinement, with 2:1 refinement ratio, centered on
the BH.  For the cases with $\tilde{\mu}=0.4$ and 0.5, we performed the
calculation at three different resolutions: the lowest has $96\times192$ points
on each refinement level and a resolution of $dx/M_0\approx 0.0256$ on the
finest level.  For the $\tilde{\mu}=0.4$ case the medium and high resolutions
have $4/3$ and $2$ times the low resolution, while for the $\tilde{\mu}=0.5$
case the medium and high resolutions are $4/3$ and $8/3$ times the low
resolution, respectively.  In Fig.~\ref{fig:econv} we show resolution studies
of the energy in the Proca field as a function of time.  This data implies
errors in measurements of the energy of the field are a few percent at the
lowest resolution, and  are less than $1\%$ for the high resolution runs
throughout the simulation run time.  We did not perform convergence studies for
the two lower $\mu$ cases, as the much slower growth rates make such studies
prohibitively expensive.  However, based on the convergence studies for the
higher $\mu$ cases, we suspect errors in the lower $\mu$ runs --- which have
larger characteristic length scales --- are still relatively small, at a few
percent at most.  In the main text all results are taken from the highest
resolution data available.
\label{conv}

\begin{figure}
\begin{center}
\includegraphics[width=\columnwidth,draft=false]{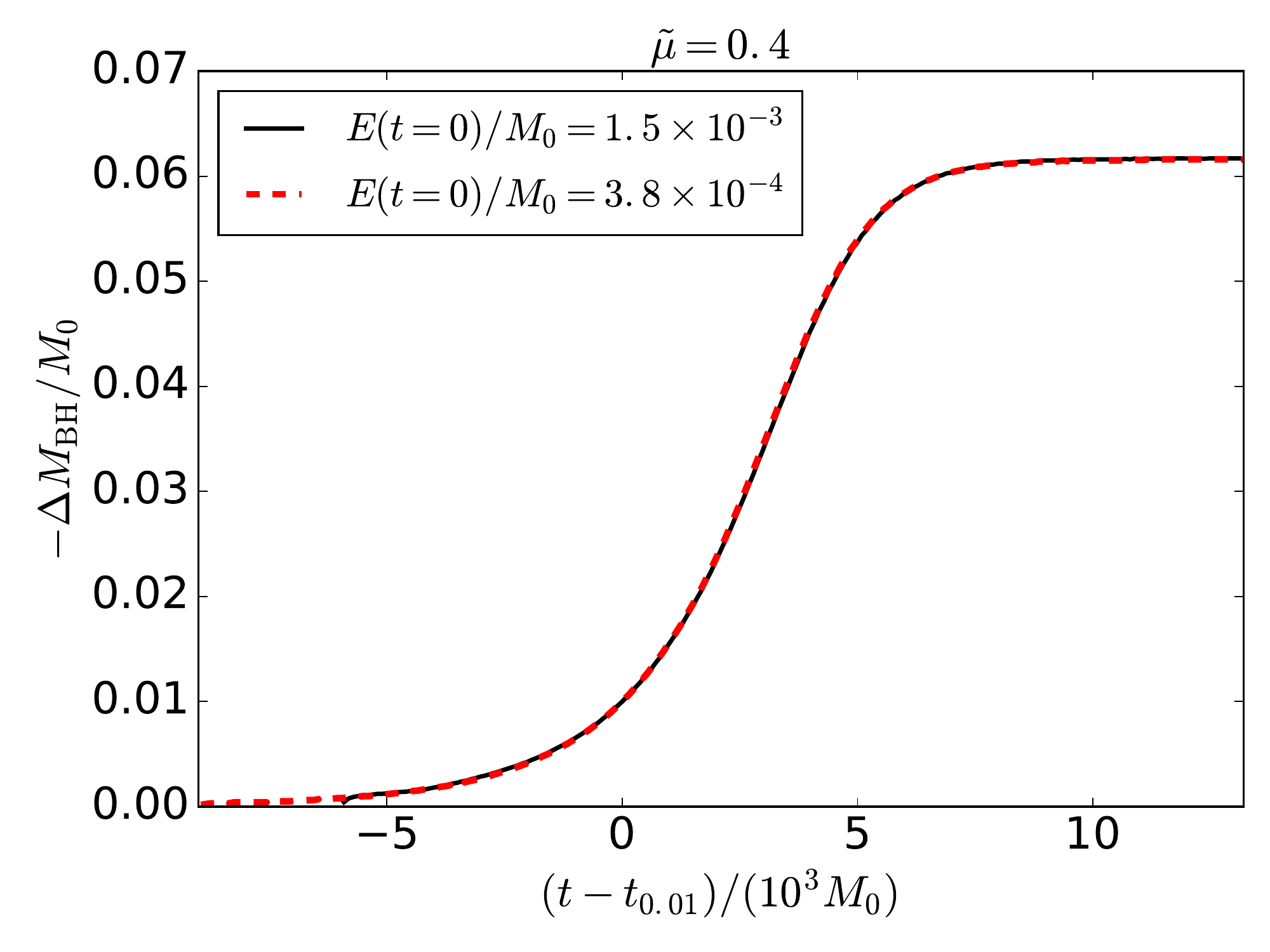}
\end{center}
\caption{
The change in BH mass as a function of time for $\tilde{\mu}=0.4$ for two 
cases: one where the initial Proca field configuration has $E(t=0)/M_0=1.5\times10^{-4}$,
and one where the initial Proca field energy is four times smaller, but otherwise identical.
A time shift has been applied so that the two curves align when $-\Delta M_{\rm BH}/M_0=0.01$. 
\label{fig:ampcomp}
}
\end{figure}

Though we do not include the effect of the perturbation of the initial Proca
field configuration on the initial BH spacetime, we have verified that it so
small as to introduce negligible error.  This is illustrated in
Fig.~\ref{fig:ampcomp} where we show a comparison of the change in BH mass as a
function of time for a simulation where we use an initial Proca field
perturbation that is half the amplitude of the standard case, for
$\tilde{\mu}=0.4$.  After applying a relative time shift (of $\approx3000$ $M_0$,
consistent with the test-field instability rate), the results from the two
cases are indistinguishable on the scale of the figure.

\end{document}